\documentclass[10pt,letterpaper]{article}

\usepackage[OE]{express}
\usepackage{amsmath,
amssymb,
graphicx,
color,
tikz,
pgfplots,
geometry,
cite,
}

\usepackage{todonotes} 

\newcommand{\ket}[1]{|#1 \rangle}

\newcommand{\ncl}{n_{\text{cl}}}

\newcommand{\micron}{~\mu\textnormal{m}}
\newcommand{\nano}{~\textnormal{nm}}

\begin{document}

\title{Digital Waveguide Adiabatic Passage Part 2: Experiment}

\author{\centering Vincent Ng$^{2,3,*}$, Jesse A. Vaitkus$^{1}$, Zachary J. Chaboyer$^{2,3}$, Thach Nguyen,$^{2,4}$, Judith M. Dawes$^{2,3}$, Michael J. Withford$^{2,3}$, Andrew D. Greentree$^{1,5}$, M. J. Steel,$^{2,3}$ }

\address{\centering $^1$Chemical and Quantum Physics, School of Science, \\ RMIT University, Melbourne 3001, Australia \\
		 $^2$ ARC Centre of Excellence for Ultrahigh bandwidth Devices for Optical Systems (CUDOS) \\
		 $^3$MQ Photonics Research Centre, Department of Physics and Astronomy, \\ Macquarie University, NSW 2109, Australia \\
		 $^4$School of Engineering, RMIT University, Melbourne 3001, Australia \\
		 $^5$Australian Research Council Centre of Excellence for Nanoscale BioPhotonics, \\ RMIT University, Melbourne 3001, Australia}

\email{\centering$^*$vincent.ng@mq.edu.au}
\begin{abstract}
Using a femtosecond laser writing technique, we fabricate and characterise three-waveguide digital adiabatic passage devices, with the central waveguide digitised into five discrete waveguidelets. Strongly asymmetric behaviour was observed, devices operated with high fidelity in the counter-intuitive scheme while strongly suppressing transmission in the intuitive. The low differential loss of the digital adiabatic passage designs potentially offers additional functionality for adiabatic passage based devices. These devices operate with a high contrast ($>\!90\%$) over a 60~nm bandwidth, centered at $\sim 823$~nm.
\end{abstract}


\bibliography{JVlibrary}

\section{Introduction}
Spatial Adiabatic Passage~\cite{ELC+2004,GCH+2004,P2006,MBA+2016} is the spatial analog of the Stimulated Raman Adiabatic Passage (STIRAP) protocol~\cite{VRS+2016}. It is a three-state transfer framework where the coupling between states is varied to effect transfer of population from one state to another, by means of an intermediate unpopulated state. This framework is remarkably flexible and resilient to variations in implementation. Adiabatic passage has numerous applications, particularly in the optical domain where it has been applied in the design of broadband optical couplers~\cite{LDO+2007,DOS+2009,CKR+2012,CCR+2012}, optical frequency conversion~\cite{PA2012} and optical photonic gates~\cite{HNM+2015}.

Recently, studies have explored adiabatic control strategies that employ digital (or piecewise) control schemes instead of continuous parameter variation~\cite{SMM+2007, SMS2009, RV2012, VG2013}. Though the condition of adiabaticity formally requires continuity, these studies have shown that adiabatic-like behaviour is maintained. These findings provide additional flexibility when designing adiabatic passage devices, particularly in cases where precise control of the coupling coefficients is difficult.

In this paper, we use the digital adiabatic passage framework and apply it to the optical waveguide domain. We fabricate digital waveguide adiabatic passage devices where the central state has been digitised into five discrete waveguidelets. 
The device designs were optimised using numerical modelling, and then fabricated using a femtosecond laser direct-write inscription technique over a range of writing powers.
These devices were characterised and shown to possess a highly transmissive configuration reminiscent of conventional adiabatic passage. Additionally in the opposite configuration, the transmission was strongly suppressed. This suppression is a key characteristic of the digitisation itself. High contrast operation ($>\!90\%$) is maintained over a 60~nm bandwidth about $\sim 823$~nm.

\section{Digital Adiabatic Passage}
The theory of digital adiabatic passage is treated in detail in \cite{VSG2016}. In its simplest form, a system consisting of three identical coupled waveguides $\{\ket{a},\ket{b},\ket{c}\}$ can be described by the Hamiltonian:
\begin{equation}
H = \left[ \begin{array}{ccc}
0 & \Omega_{ab} & 0 \\
\Omega_{ab} & 0 & \Omega_{bc} \\
0 & \Omega_{bc} & 0 
\end{array} \right],
\qquad
\ket{E_0} = \frac{\Omega_{bc} \ket{a} - \Omega_{ab} \ket{c}}{\sqrt{\Omega_{ab}^2+\Omega_{bc}^2}}. 
\end{equation}
where the eigenstate $\ket{E_0}$ is the so-called dark-state of $H$. Here, $\Omega_{nm}$ is the coupling between the $n^{\text{th}}$ and $m^{\text{th}}$ waveguides, and only nearest neighbour coupling has been assumed. 

This eigenstate $\ket{E_0}$ is completely composed of states $\ket{a}$ and $\ket{c}$. The process of adiabatic passage corresponds to the use of this eigenstate: starting with the coupling coefficients such that $\Omega_{bc}  \gg \Omega_{ab}$ and slowly varying them such that $\Omega_{ab} \gg \Omega_{bc}$ at the end of the device, transport is effected from $\ket{a}$ to $\ket{c}$. Due to the ordering of the coupling, this is commonly called the counter-intuitive coupling scheme. When the coupling order is reversed (corresponding to launching light in $\ket{c}$ for the same coupling order), this is known as the intuitive coupling scheme, and leads to non-adiabatic oscillations, although, with central state detuning can lead to bright-state adiabatic passage~\cite{GNH+2009}. As the name suggests, digital adiabatic passage devices are realised by digitising the central waveguide of standard waveguide adiabatic passage into several parallel elements which we term \emph{waveguidelets}. For ideal systems with equal propagation terms or no direct next nearest neighbour ($a$--$c$) coupling, the \textit{effective} $a$--$c$ hopping rate~\cite{VG2013} dictates the ideal segment length: $L_{\text{opt}} = \pi/\sqrt{\Omega_{ab}^2 + \Omega_{bc}^2}$.


\section{Model Parameters}
To simulate the device, model parameters were heuristically obtained from previous work by \cite{CMH+2015}. For a detailed discussion of the digital adiabatic passage device see \cite{VSG2016}. Both model parameters describing the refractive index difference $\delta$, and the $1/e$ width $\rho$ of the waveguides were varied in the software package BeamPROP until they gave a suitably good fit to the experimentally obtained mode field diameters and coupling curves. 
Rigorous 3D field propagation simulation was then performed to obtain the coupling lengths. This simulation was carried out using an in-house developed propagation tool based on the eigenmode expansion method \cite{NTK+2009}. Optimised lengths were found by inferring the coupling between guides from the beat length of two-waveguide systems. Although the eigenmode expansion is highly efficient, it is not able to model scattering losses.  Accordingly, once a suitable design was chosen, the geometry was input into BeamPROP for accurate analysis of the scattering. Device parameters used in simulations can be found in table \ref{tab:device2}.

\begin{table}[h]
\centering
\caption{Device parameters used in all calculations herein. All waveguidelets are aligned at $y=0$ and $\ket{a}, \ket{b}_1, \ket{c}$ all begin at $z=0$. All waveguidelet pairs $\ket{b}_{i+1}$ and $\ket{b}_{i}$ are separated in $z$ by 7.5~mm to increase the total length to 70 ~mm to further demonstrate digitisation.}
\label{tab:device2}
\begin{tabular}{cccccccc}\hline
Waveguidelet & $\ket{a}$ & $\ket{b}_1$ & $\ket{b}_2 $ & $\ket{b}_3$ & $\ket{b}_4$ & $\ket{b}_5$ & $\ket{c}$ \\ \hline
$L_{\text{opt}} $(mm) & N/A & 5.5824 & 9.2295 & 10.3775 & 9.2295 & 5.5824 & N/A \\ 
$x$ ($\mu$m) & -10.00  & -2.00 & -0.75 & 0.00 & 0.75 & 2.00 & 10.00  \\ \hline
\end{tabular}\vspace{0.5em}
\begin{tabular}{cccccccc}\hline
$\rho$ & 1.6 $\mu$m & $\ncl$ & $1.4994$ & $\delta$ & $0.0056$  & $\lambda_{\text{opt}}$ & 800 nm \\ \hline
\end{tabular}
\end{table}

\begin{figure}[h]
\centering
\includegraphics[width=0.47\textwidth]{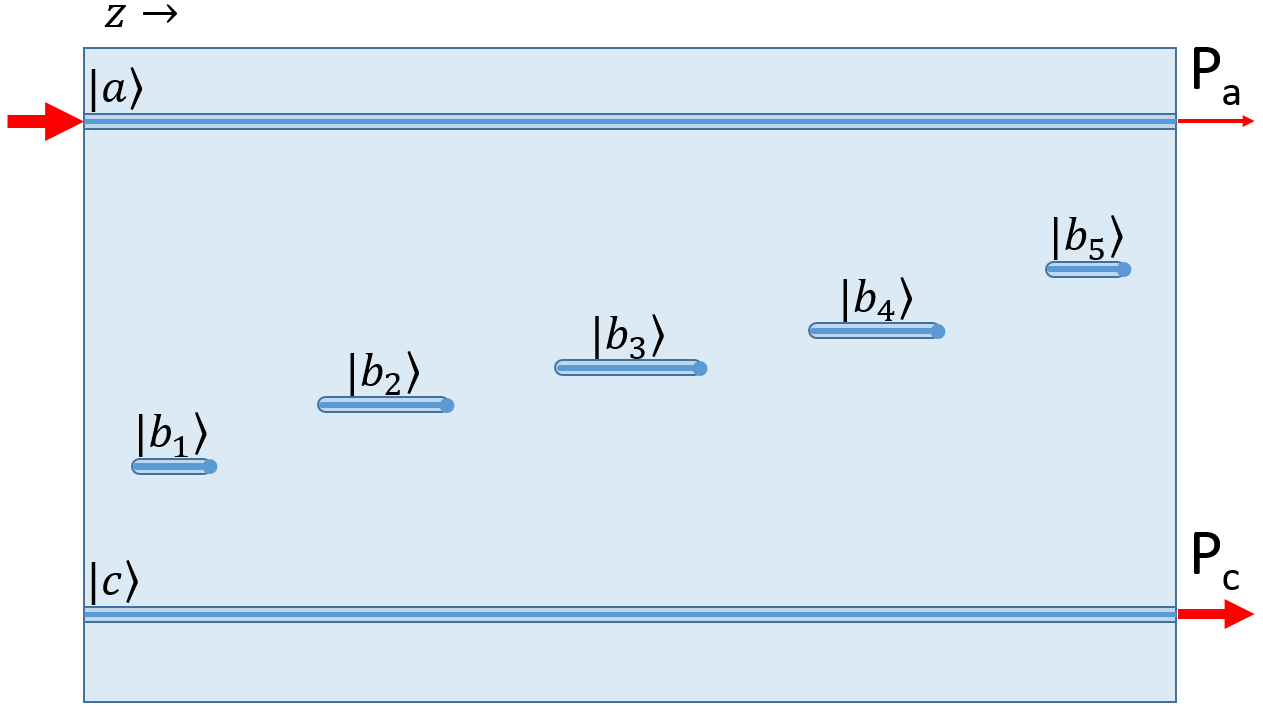}
\includegraphics[width=0.47\textwidth]{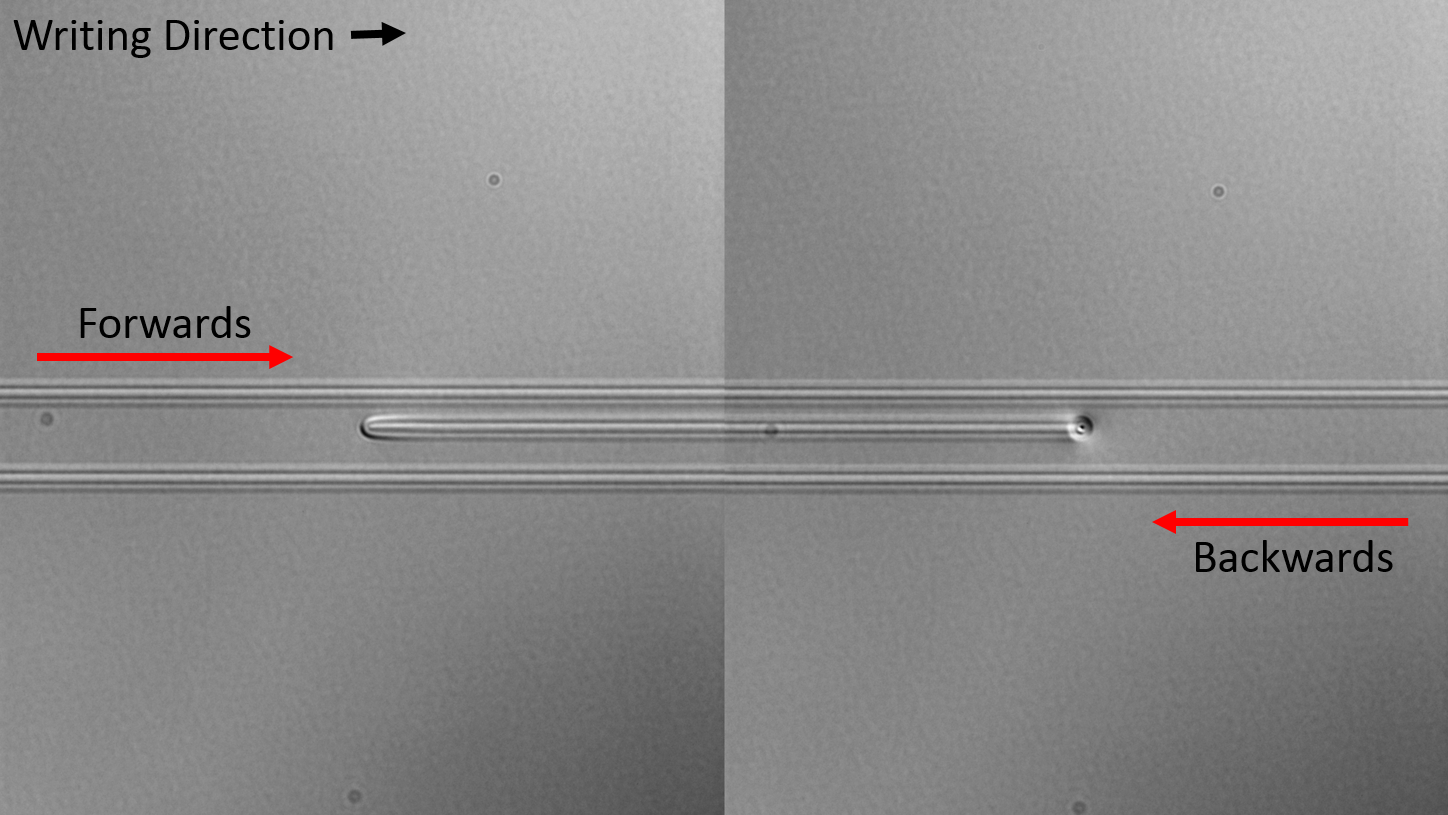}
\caption{(Left) Schematic of device with parameters determined by Table \ref{tab:device2}. (Right) Stitched differential interference contrast (DIC) microscope image of the start and end of a waveguidelet making up the dark state. A taper and a void are evident at either end. 
}
\label{design}
\end{figure}

\section{Implementation}
The waveguides in all devices herein were fabricated using an ultrafast laser inscription technique~\cite{DMS+1996}. The output of a Ti: Sapphire oscillator (Femtolasers GmbH, FEMTOSOURCE XL 500, 800 nm centre wavelength, 5.1 MHz repetition rate, $<$50 fs pulse duration) was focused into a borosilicate substrate (Corning Eagle2000) at a depth of 170~$\mu$m using a 100x oil immersion objective lens. The sample was translated with respect to the beam focus using Aerotech motion control stages with 10 nm precision. Our combination of writing parameters lies within the cumulative heating regime of refractive index modification~\cite{SBC+2001} in which consecutive pulses are absorbed within the focal volume before the dissipation of energy into the bulk of the material, leading to a refractive index modification dominated by thermal effects. The thermal mechanism of the refractive index modification causes both the peak contrast and physical size of the waveguide to increase with the amount of absorbed energy~\cite{EZN+2008}. This allowed the index profile of the written waveguides to be controlled by varying the writing pulse energy. The writing pulse energy was iterated between 28.5 and 34~nJ in 0.25~nJ steps at a constant sample feedrate of 1500~mm/min to obtain waveguides with a refractive index profile matching our design for optimal operation at $\lambda = 800$~nm. 

The design of the fabricated adiabatic passage devices is shown in Fig.~\ref{design}(a). The total device length including spaces is 70~mm. The input and output states $\ket{a}$ and $\ket{c}$ consist of straight waveguides spaced by 20$\micron$. This choice was made so that there would be negligible direct coupling between the outer waveguides. The central waveguide $\ket{b}$ was digitised into 5 waveguidelets that were written by modulating the laser output with a fast RTP pockels cell. A taper is observed at the start of the waveguidelet, as the threshold for index modification, due to cumulative heating, is gradually reached after illumination by the laser pulses. On the other hand, the sudden turning-off of the laser at the waveguidelet end leads to a rapid transition away from the thermal regime and the formation of a bulbous void \cite{MWF2011} as shown in Fig.~\ref{design}(b). These features are expected to introduce asymmetric losses in the device for the forward and backward launch directions, indicated in Fig.~\ref{design}(b). 

The scattering losses are also expected to be asymmetric with respect to the intuitive and counter-intuitive configurations.  
The scattering is expected to be weak in the counter-intuitive configuration due to the dark-state. Conversely in the intuitive configuration, the central state is populated and the voids should scatter strongly, leading to a differential loss in the devices.
Each device was accompanied by a reference waveguide for transmission measurements and two waveguides with a 20$\micron$ spacing to verify that direct coupling between states $\ket{a}$ and $\ket{c}$ is negligible. 

\section{Characterisation}
Eleven of the devices fabricated are characterised below, written with writing powers between 28.75--31.5~nJ per pulse. These devices yielded coupling coefficients most consistent with modelling. 
In order to verify adiabatic passage behaviour in these devices, light at 808~nm was fibre coupled into the chip in the counter-intuitive configuration and the outputs measured.

\begin{figure}[h]
\centering
\includegraphics[width=0.32\textwidth]{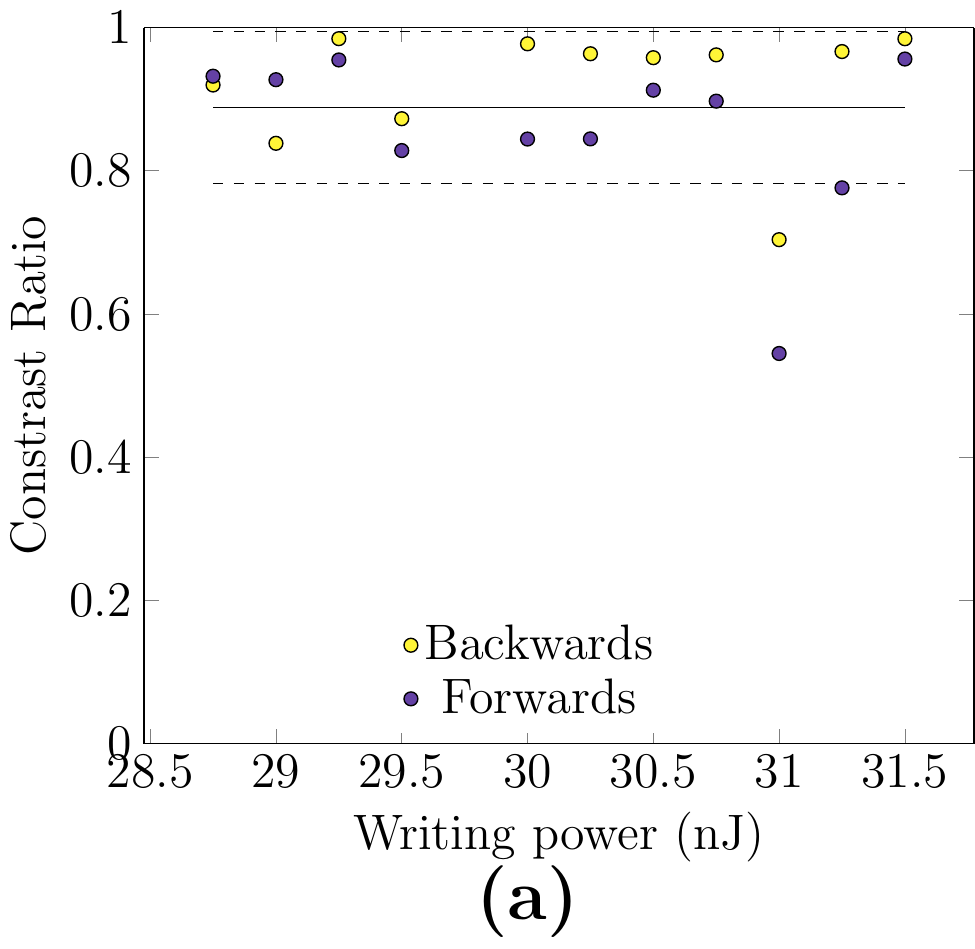}
\includegraphics[width=0.32\textwidth]{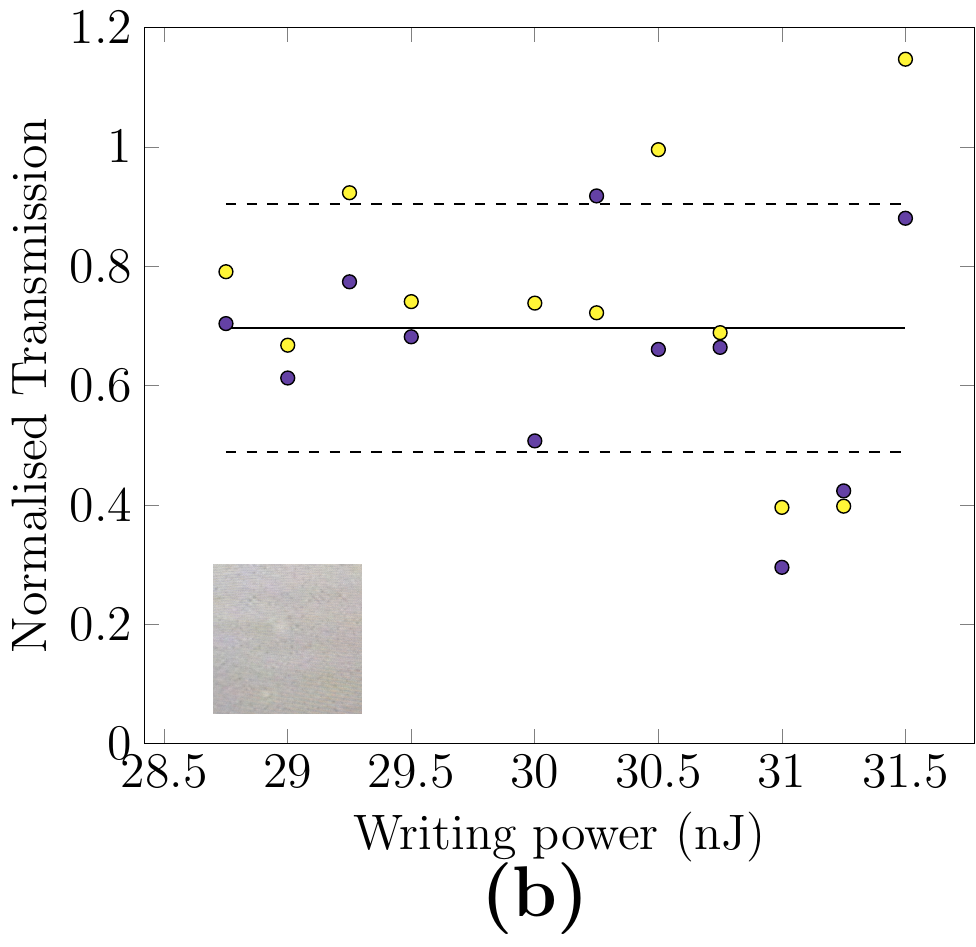}
\includegraphics[width=0.32\textwidth]{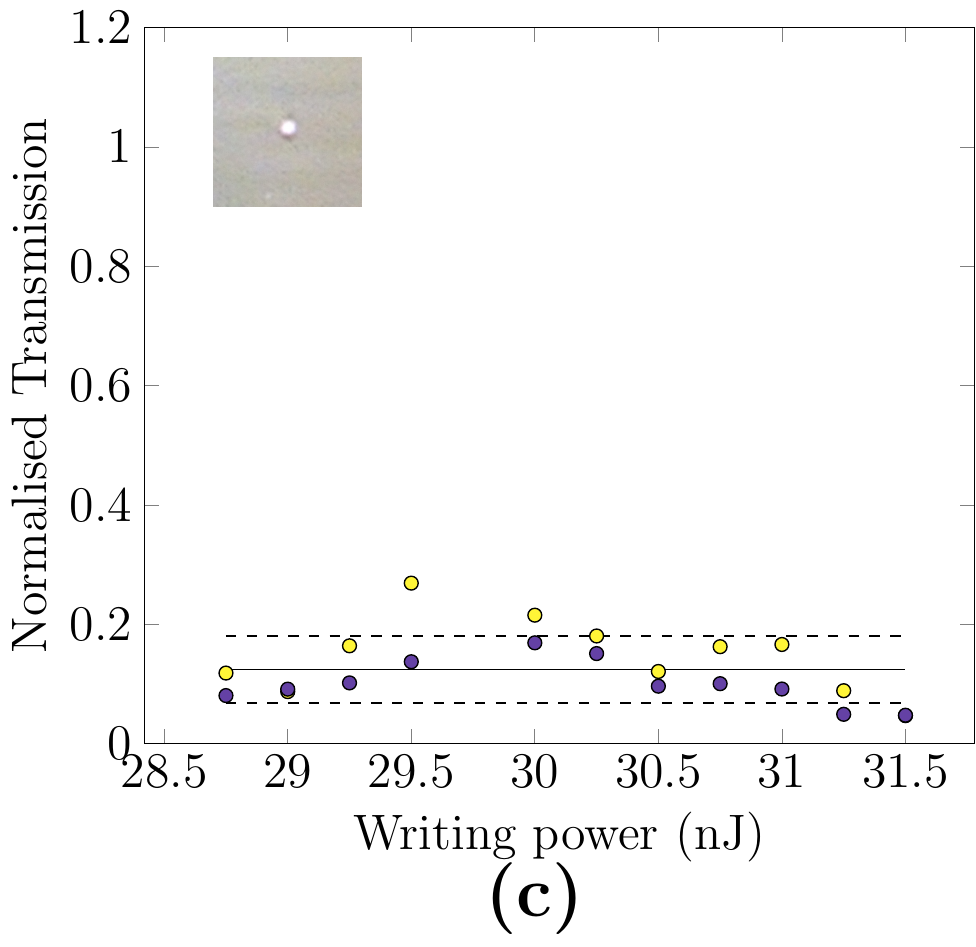}
\caption{Characterisation at 808~nm. The contrast ratio is plotted as $P_c/(P_a+P_c)$ for the counter-intuitive configuration (a). The total transmission is also plotted as $(P_a+P_c)/P_{ref}$, for the counter-intuitive configuration (b) and the intuitive configuration (c). Insets show a CCD image of the voids at the end of the waveguidelets. The voids are shown to be bright and strongly scattering in the intuitive configuration, but dark in the counter-intuitive configuration where they remain largely unpopulated. Transmission is higher in the backwards direction in almost all cases. Mean (solid line) and standard deviation (dashed) have been indicated.}
\label{vis}
\end{figure}

The contrast ratio between the waveguide outputs is plotted in fig. \ref{vis}(a) as $P_c/(P_a+P_c)$, for each device in the counter-intuitive configuration. Here, $P_a$ and $P_c$ are the output powers of waveguides $\ket{a}$ and $\ket{c}$ respectively. The 11 devices tested showed little-to-no dependence on the writing energy over this range, and light was observed to have coupled across the device in the counter-intuitive configuration in almost all cases. These demonstrate insensitivity of the devices to the effective device length, characteristic of adiabatic passage designs.
The largest source of variability is believed to be waveguide inhomogeneity due to the relatively large length of the devices. On average, $89\%$ of the output is successfully coupled. Asymmetry in fidelity is also evident in the forwards and backwards directions, with a slightly greater fidelity in the backwards direction. This is easily explained by the position of the voids---in the backwards direction, each void is present at the beginning of the waveguidelet, where the light has yet to couple across and be scattered.


The outputs of both waveguides were also summed and normalised against a straight waveguide written with the same power, as shown in fig. \ref{vis}(b). 
The losses can mainly be attributed to scattering arising from digitisation, with variation from device to device arising from waveguide defects. Despite this, the average device yields a $70\%$ transmission.
In contrast, light launched in the intuitive configuration  is consistently suppressed, transmitting just $12\%$ of the light. Without the dark state in the intuitive direction the waveguidelets become strongly scattering, as shown in the inset of fig. \ref{vis}(c). A slight difference in the forwards and backwards directions is also evident, with greater losses in the forwards direction for reasons discussed earlier in the paper.

\begin{figure}[h]
\centering
\includegraphics[width=0.47\textwidth]{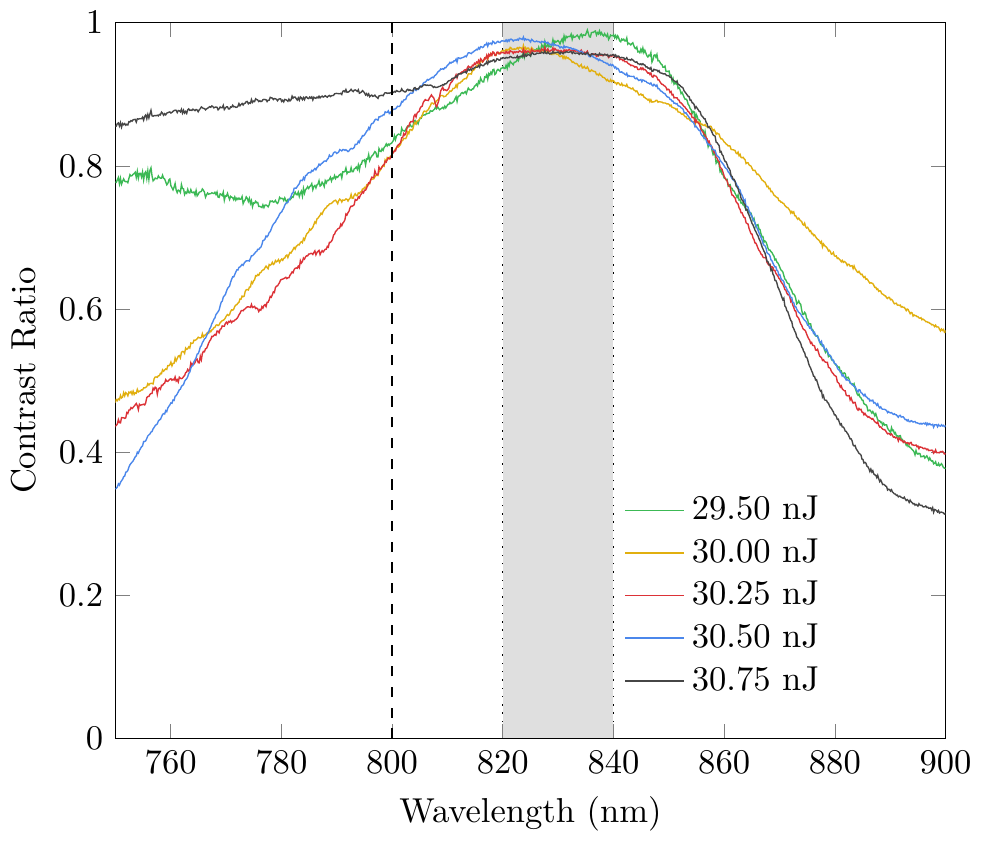}
\qquad
\includegraphics[width=0.47\textwidth]{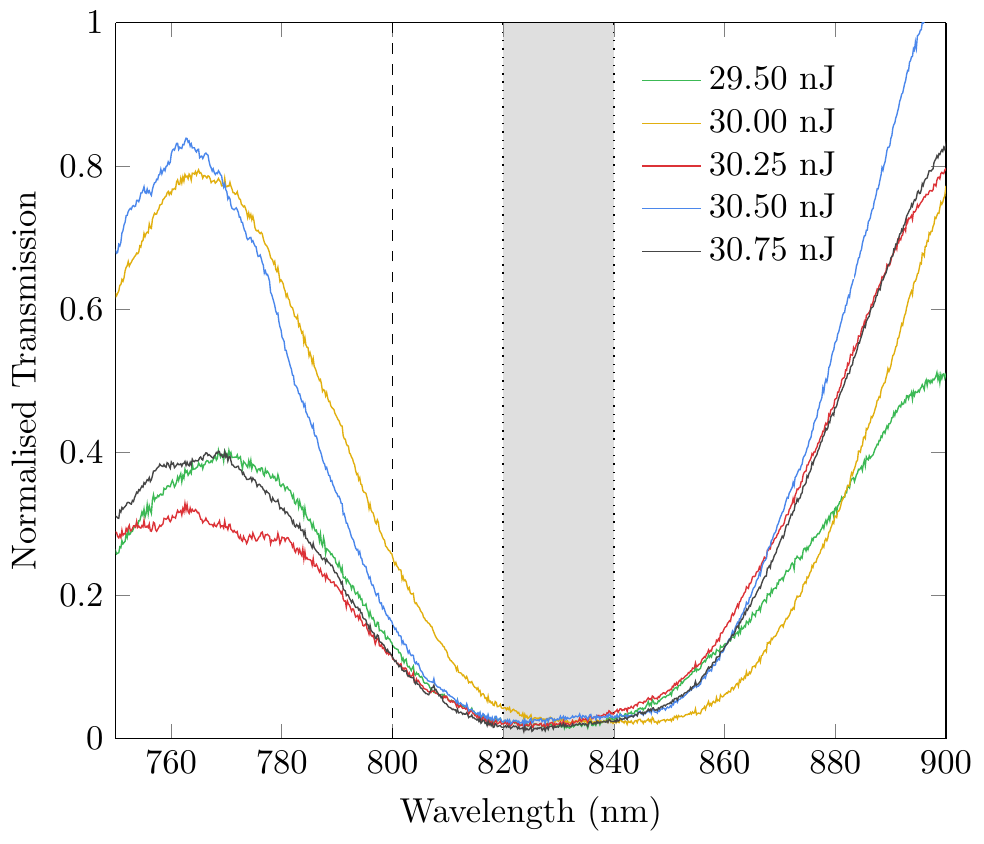}
\caption{Spectral response of the devices are plotted, measured in the forwards direction. Counter-intuitive contrast ratio (left) plotted as $P_c/(P_a+P_c)$, and Intuitive transmission (right) is plotted as $(P_a+P_c)/P_{\text{ref}}$. Note that the optimal operating wavelength is shifted from 800~nm (dashed) to be about 830~nm (shaded). It can be seen that the broad wavelength response in the counter-intuitive configuration coincides with a suppressed response in the intuitive configuration.}
\label{spectra}
\end{figure}

To test the operating bandwidth, the devices were also characterised using a white light source (NKT Photonics SuperK Compact). The white light spectrum was narrowed down using a FGB25 colour filter to capture the spectral response of the devices between 700~nm and 900~nm. The output was fibre-coupled to a USB4000 OceanOptics spectrometer. The devices shown were measured in the forwards direction.

The wavelength dependent performance of these devices is shown in fig. \ref{spectra}(a). These were normalised against straight reference waveguides in order to remove the spectral shaping caused by the filtering and the white light source itself. Across the writing powers used, significant variability can be observed at the short and long wavelengths, but a region where a high contrast ratio is maintained is consistently observed in a wavelength band centered around $\sim 830$~nm. This is shifted from the intended centre wavelength of 800~nm.   Despite this shift and the device variability, each device exhibits fidelity above $95\%$ at its optimal wavelength.
The devices also remain insensitive to small variations in writing power, reflected in the small shifts in optimal frequency. The optimal digital adiabatic passage performance coincides with a  suppressed transmission in the intuitive configuration, shown in fig. \ref{spectra}(b), with a throughput of $<5\%$.

\begin{figure}[h]
\centering
\includegraphics[width=0.47\textwidth]{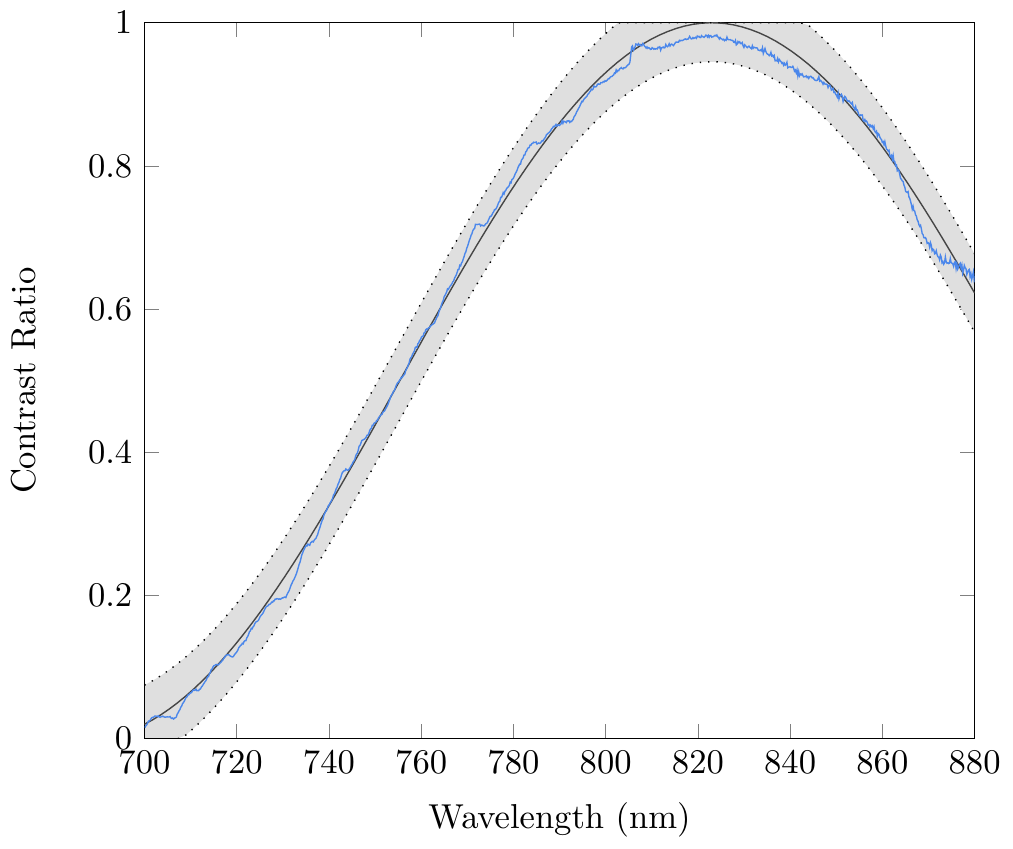}
\qquad
\includegraphics[width=0.47\textwidth]{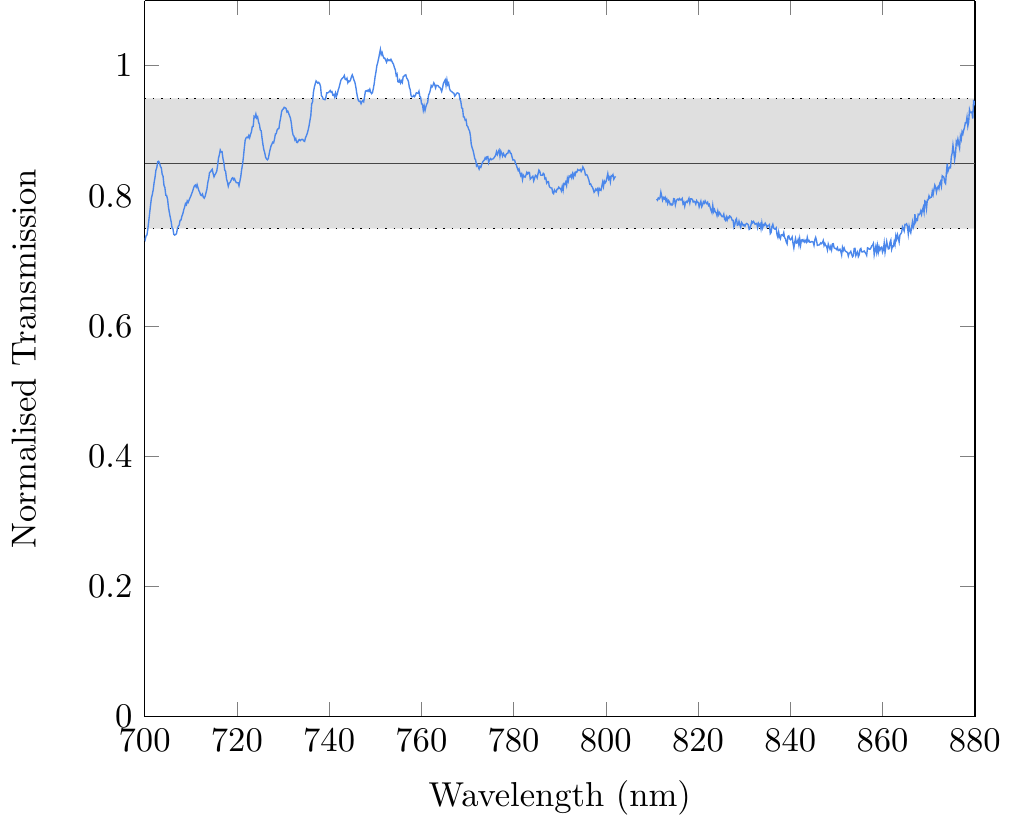}
\caption{Second set of measurements taken of waveguide written at 30.5nJ, using bandpass filtering. (left) Contrast ratio $P_a/(P_a+P_c)$ as a function of wavelength. Shaded area corresponds to 95\% confidence interval using bisquare method. The fitted equation is $\cos^2(2 \pi (\lambda-\lambda_{opt})/\lambda_{\lambda})$ where $\lambda_{opt} = 823.00 \pm 0.17$~nm and $\lambda_{\lambda} = 541.15 \pm 1.39$~nm. (right), Device transmission $(P_a+P_c)/P_{\text{ref}}$ as a function of wavelength. A section of the transmission around 805nm has been omitted due to a normalisation error. In the optimum region, transmission typically lies between 75\% to 95\% (shaded in-image)}
\label{D9}
\end{figure}

The operational bandwidth with a fidelity above $90\%$ is $\approx 60\nano$, after which the adiabatic passage-like behaviour rolls off on either side. Switching the filter from the FGB25 to a set of long-pass and short pass filters allowed us to shift the spectral window and observe the smooth roll off of the 30.5~nJ device shown in fig. \ref{D9}. This roll off is consistent with modelling \cite{VSG2016}.
The total transmission of this device is also plotted, where the transmission around 805~nm was omitted due to a normalisation error arising from the saturation of the spectrometer. Wavelength dependent loss is evident in this device, and is assumed to come from the scattering properties of the voids terminating the waveguidelets. On average, more than $80\%$ of the light was transmitted across the spectrum tested.

The bandwidth of these devices, constrained by the effective a-c hopping rate and the waveguidelet lengths, was measured to be narrower than predicted in \cite{VSG2016}. Both this and the shift in the center wavelength suggests either a shift in the profile height parameter $\Delta$ or the width $\rho$, or that the Gaussian graded index model is an invalid approximation in estimating mode coupling. In principle, the bandwidth can be improved by increasing the number of waveguidelets, reducing the digitisation error.

\section{Conclusions}
We have experimentally demonstrated digital adiabatic passage devices, using the design constraints  investigated in \cite{VSG2016}, and fabricated using a femtosecond laser direct write technique. 
Despite the variability from device to device, as well as the wavelength dependent behaviour, the devices still strongly exhibit features characteristic of adiabatic passage. These features include robustness against variations in coupling, robustness against strong scattering losses in the central state, $\ket{b}$ (from both the digitisation and writing asymmetry), in the counter-intuitive configuration. Additionally, as a consequence of the digitisation process, these features are consistently accompanied by a suppression of the transmission the intuitive configuration, providing a new functionality for digital adiabatic passage devices. These characteristics suggest digital adiabatic passage may be a robust framework for designing photonic devices with novel applications. 

\section*{Acknowledgements}
This research was supported by the ARC Centre of Excellence for Ultrahigh bandwidth Devices for Optical Systems (Project Number CE110001018) and was performed in part at the OptoFab node of the Australian National Fabrication Facility using NCRIS and NSW state government funding. A.D.G. acknowledges the ARC for financial support (Grant No. DP130104381).

\end{document}